\documentclass[letter]{aa}
\pdfoutput=1

\usepackage{graphicx}
\usepackage{txfonts}
\usepackage{pictex}
\usepackage{latexsym}
\usepackage{graphics}
\usepackage{afterpage}
\usepackage{units}

\begin{document} 

   \title{Gaia-supported re-discovery of a remarkable weak line quasar  from a variability and proper motion survey\thanks{Based on observations with the 2.2 m telescope of the German-Spanish Astronomical Center, Calar Alto, jointly operated by the Max-Planck-Institut f\"ur Astronomie Heidelberg and the Instituto de Astrof\'isica de Andaluc\'ia (CSIC).}}

   \author{Helmut Meusinger\inst{1}
          and
           Ralf-Dieter Scholz\inst{2}
          }

   \institute{Th\"uringer Landessternwarte, Sternwarte 5, D-07778 Tautenburg, Germany 
                   \email{meus@tls-tautenburg.de}
                   \and
                   Leibniz-Institut für Astrophysik Potsdam (AIP), An der Sternwarte 16, D-14482 Potsdam, Germany
                 }

  \date{Received June 28, 2022; accepted July 28, 2022}

 \abstract{
We demonstrate that VPMS\,J170850.95+433223.7  is a weak line quasar (WLQ) which is remarkable in several respects.
It was already classified as a probable quasar two decades ago, but with considerable uncertainty. 
The non-significant proper motion and parallax from the Gaia  early data release 3  have solidified this assumption.  
Based on previously unpublished spectra, we show that VPMS\,J170850.95+433223.7  is a WLQ at $z = 2.345$ with immeasurably faint broad emission lines in the rest-frame ultraviolet. 
A preliminary estimate suggests that it hosts a supermassive black hole of $\sim 10^9\,M_\odot$ accreting close to the Eddington limit, perhaps  at the super-Eddington level. 
We identify two absorber systems with blueward velocity offsets of $0.05c$ and $0.1c$, which could represent high-velocity outflows, which are perhaps related to the high accretion state of the quasar.
 }

% 5 {} token are mandatory


  \titlerunning{Remarkable weak line quasar}
  \authorrunning{H. Meusinger and R. Scholz}

   \maketitle
%
%

%**********************************************************************************
%
\section{Introduction}\label{Intro}
%
%**********************************************************************************

The weakness of broad emission lines is the defining characteristics of a rare class of high-luminosity active galactic nuclei (AGN) called weak-line quasars (WLQs).
They were first discovered about a quarter of a century ago \citep{McDowell_1995, Fan_1999}. 
The Sloan Digital Sky Survey \citep[SDSS;][]{York_2000} has revealed the existence of a (still relatively small) population of WLQs 
\citep[e.g.][]{Diamond_2009, Plotkin_2010, Meusinger_2014}.
\footnote{Following \citet{Diamond_2009},  WLQs are defined as quasars having equivalent widths (EWs) of either  the Ly$\alpha$+\ion{N}{v} emission line complex or the \ion{C}{iv} line  below the $3\sigma$ threshold  of the EW distribution in the parent quasar sample, that is $\mbox{EW(Ly$\alpha$+\ion{N}{v})} < 15.4$\,\AA\ \citep{Diamond_2009} and  $\mbox{EW(\ion{C}{iv})} < 4.8$\,\AA\ \citep{Meusinger_2014}.}
A number of different scenarios have been proposed, including, in particular, extremely high accretion rates, anemic broad emission line regions, and a gravitational lensed accretion disk (AD).
Currently, the idea of a shielding gas component between the central X-ray source and the broad emission line region appears particularly attractive
\citep[e.g.][and references therein]{Paul_2022}.

The Tautenburg -- Calar Alto Variability and (zero) Proper Motion Survey (VPMS) is a quasar search project that is based on optical long-term variability and non-detectable proper motions measured on a large number of imaging observations with the Schmidt camera of the Tautenburg 2-m telescope \citep[][]{Scholz_1997, Meusinger_2001, Meusinger_2002}.  
One of the goals of VPMS was to search for quasars with odd spectra that might not be picked out in colour-based quasar surveys. 
An example of this is the unusual broad-absorption line (BAL) quasar VPMS\,J134246.24+284027.5 \citep{Meusinger_2005}. 
Here, we present another remarkable quasar discovered from this survey, the WLQ \object{VPMS\,J170850.95+433223.7} (hereafter VPMS\,J1708+4332). 
Both sources are located in the SDSS footprint area, but were not targeted for spectroscopy by SDSS. 

VPMS\,J1708+4332 was classified as a high-priority quasar candidate in the VPMS field around the globular cluster M~92.  
Follow-up observations yielded a spectrum similar to that of a WLQ, but at first it could not be assigned with reasonable certainty to a quasar.
It was not detected as a radio source at the flux level of  the FIRST survey \citep{Becker_1995}. 
Recently, the Gaia early data release 3 \citep[EDR3;][]{Gaia_2021_1G} has provided data that make a significant contribution to distinguishing between quasars and foreground stars in the case of unclear spectra. 

We present and interpret the spectra and the Gaia data in Sect.\,\ref{sect:Obs}.  
In Sect.\,\ref{sect:Discussion}, we discuss some properties of the quasar, Sect.\,\ref{sect:Summary} gives the conclusions.
We assume  Lambda Cold Dark Matter ($\Lambda$CDM) cosmology with $H_0 = 73$ km s$^{-1}$ Mpc$^ {-1}$, $\Omega_{\rm \Lambda} =0.73$, and $\Omega_{\rm M}=0.27$.

%**********************************************************************************
%
\section{Observations and analysis of the spectra}\label{sect:Obs}
%
%**********************************************************************************

VPMS\,J1708+4332 was observed with the focal reducer and faint object spectrograph CAFOS at the 2.2-m
telescope of the German-Spanish Astronomical Centre (DSAZ) on Calar Alto, Spain, in July 1998. 
The grism B-400 was used with a wavelength coverage of $3200 - 8000$\,\AA\ and a dispersion of 10\,\AA\,px$^{-1}$  on the
SITe1d CCD.  In a subsequent campaign in July 2004,  VPMS\,J1708+4332 was observed at higher resolution using 
the grisms B-200 and B-100, which have a wavelength coverage of  $3200 - 7000$\,\AA\ and $3200 - 5800$\,\AA\, and a dispersion of $\sim 5$\,\AA\,px$^{-1}$ and $\sim 2$\,\AA\,px$^{-1}$, respectively. One B-200 spectrum and two B-100 spectra were recorded, each with an exposure time of 1200 seconds.
The spectra were reduced using the ESO MIDAS data reduction package with standard procedures including bias correction, flat-fielding, cosmic ray removal, 
sky subtraction, wavelength calibration, and a rough flux calibration.  
The wavelength calibration was done by means of calibration lamp spectra. 
An exact flux calibration was not carried out because it was not considered absolutely necessary at that time. 
Galactic foreground reddening was corrected adopting $E(B-V) = 0.014$ from \citet{Schlafly_2011} and the Milky Way reddening curve from \citet{Pei_1992}.  

\begin{figure}[htbp]
\begin{center}
\includegraphics[viewport= 20 20 590 770,width=6.5cm,angle=270]{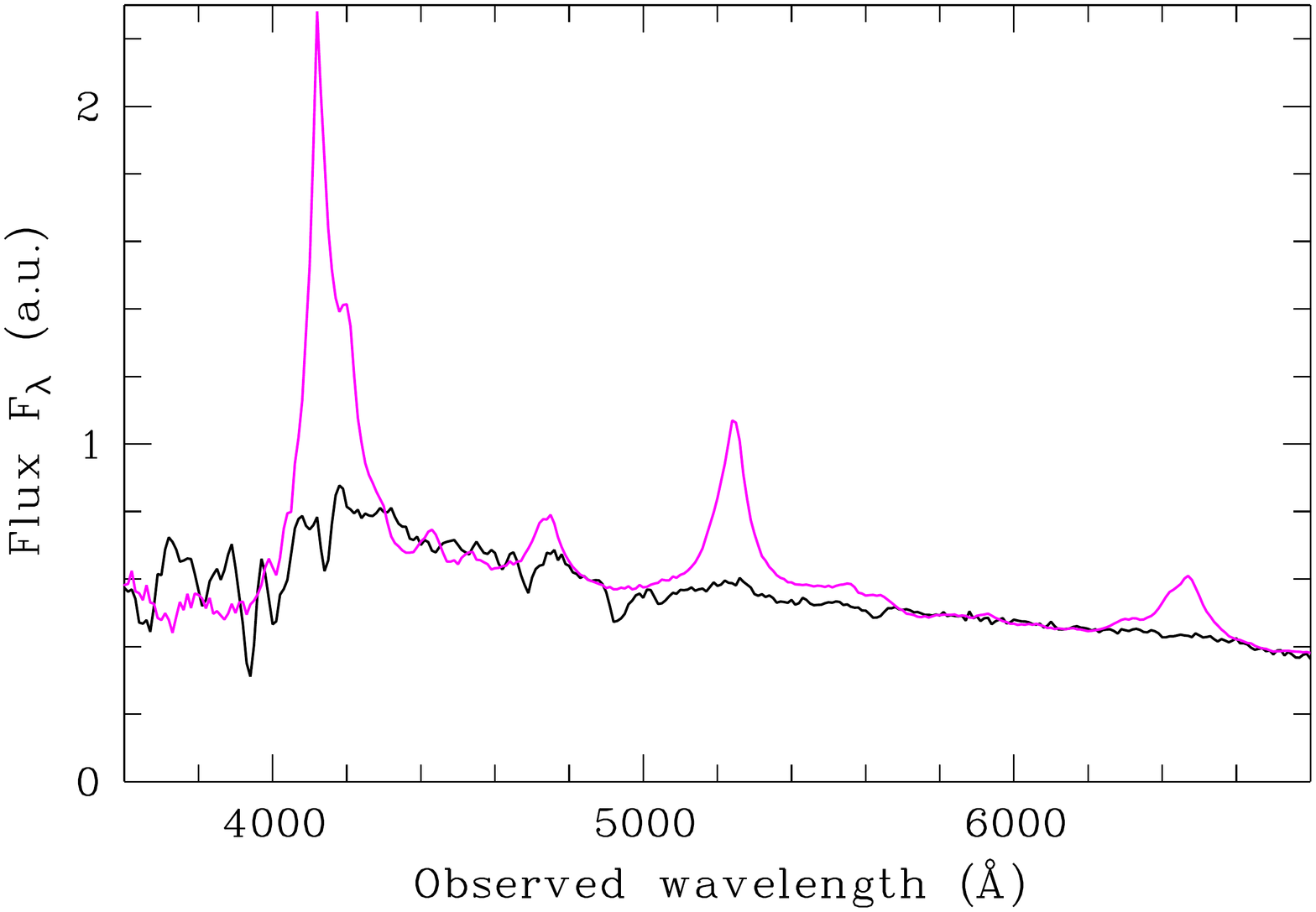} 
\caption{CAFOS B-400 spectrum of VPMS\,J1708+4332 (black) in arbitrary units. 
For comparison, the arbitrarily scaled SDSS quasar composite spectrum \citep{Vanden_Berk_2001} is shown (magenta) 
redshifted into the observer frame with $z = 2.39$ \citep{Meusinger_2001}. 
}
\label{fig:B400}
\end{center}
\end{figure}

In a first attempt \citep{Meusinger_2001}, we had identified the peak in the B-400 spectrum of VPMS\,J1708+4332 with the Ly\,$\alpha$ + \ion{N}{v} line complex and thus estimated a redshift of $z \approx 2.39$ from the best fit of the SDSS quasar composite spectrum template \citep{Vanden_Berk_2001} to the continuum (Fig.\,\ref{fig:B400}). 
Later, \citet{Richards_2009} listed this source as a quasar candidate with a photometric redshift  $z_{\rm ph} = 2.215$ based on the SDSS magnitudes. 
The mid-infrared (mid-IR) colour indices $W1-W2=0.8$ and $W2-W3=3.7$ from the Wide-field Infrared Survey Explorer \citep[WISE;][] {Wright_2010} are typical for quasars \citep{Jarrett_2017} and thus also seem to support this assessment.
But, nevertheless, the classification of VPMS\,J1708+4332 as a quasar could not be considered to be certain especially because the absorption features in our spectrum could not be convincingly explained.

We therefore also considered the alternative possibility that  VPMS\,J1708+4332  might be a rare star type.
If so, the spectrum most closely resembles that of an O-type subdwarf \citep{Jeffery_2021}:
A relatively strong line is seen close to the position of \ion{He}{ii}\,$\lambda\,4686$, which, however, turns out to be a double line in the B-100 spectrum.
There are no hydrogen lines, perhaps with the exception of a weak feature at 4863\,\AA, where hot subdwarfs show a blend of H$\beta$ with a \ion{He}{i} line.  
According to the scheme of \citet{Jeffery_2021}, which is based on line ratios, we would have to classify the spectrum as sdO3-4. 
Such stars are very hot  ($T_{\rm eff} \ga 30\,000$\,K). In the optical and near-IR, the spectral energy distribution (SED)  can be fitted by a black body of such a high temperature only if strong reddening is invoked, for instance  $E(B-V) = 0.75$ for a 35\,000\,K black body. 
The source of the reddening could be related to the dusty component which is indicated by  the above-mentioned mid-IR excess.
The distance to such a strongly reddened sdO star is estimated to  be $\sim 3$\,kpc, which raises the question of whether this is compatible with 
the zero-proper motion from VPMS.

\begin{figure}[htbp]
\begin{center}
\includegraphics[viewport= 0 0 630 440,width=9.0cm,angle=0]{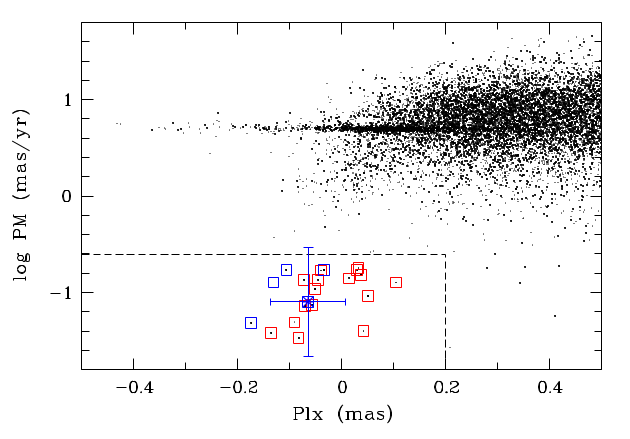} 
\caption{Proper motion (PM) vs. parallax (Plx) diagram from Gaia EDR3 for the objects with parallax errors  $<0.1$ \,mas
within two degrees around VPMS\,J1708+4332.
Red frames, quasars from SDSS DR16 \citep{Ahumada_2020}; blue frames, quasar candidates from \citet{Richards_2009};
and blue asterisk, VPMS\,J1708+4332. For clarity, error bars have only been plotted for VPMS\,J1708+4332.
The concentration of sources at log\,PM $\sim 0.75$ is due to the  globular cluster M~92.
}
\label{fig:PM_Plx}
\end{center}
\end{figure}

With the availability of the data from the Gaia satellite, the question of zero-proper motion (and parallax) can be answered on a more secure basis.
Gaia EDR3 \citep[][]{Gaia_2021_1G} lists a non-significant, slightly negative parallax $Plx = -0.0648  \pm 0.0725$\,mas
and a non-significant, zero proper motion $pmRA = -0.076 \pm 0.082$\,mas/yr, $pmDE = +0.023 \pm 0.090$\,mas/yr.  
Zero parallaxes and proper motions are exceptional among the objects with $Plx < 0.5$\,mas and parallax errors $< 0.1$\,mas in the sky region around VPMS\,J1708+4332 and only shown by known  quasars  or quasar candidates (see Fig.~\ref{fig:PM_Plx}). 
Considering objects with similar magnitudes  ($Gmag = 17.555 \pm 0.5$\,mag) within 20\,arcmin from VPMS\,J1708+4332, we compared its various EDR3 astrometric quality parameters, for example parallax and proper motion errors, 
\textsf{astrometric\_sigma5d\_max},
\textsf{astrometric\_excess\_noise},
\textsf{ipd\_frac\_multi\_peak}, 
\textsf{ruwe},
\textsf{astrometric\_gof\_al}, and
\textsf{visibility\_periods\_used} 
\citep[cf. Table A.1 in][]{Gaia_2021_6G},
with their typical regional values and found no obvious outliers.

\begin{figure*}[htbp]
\begin{center}
\includegraphics[viewport= 0 0 324 720,width=7.8cm,angle=270]{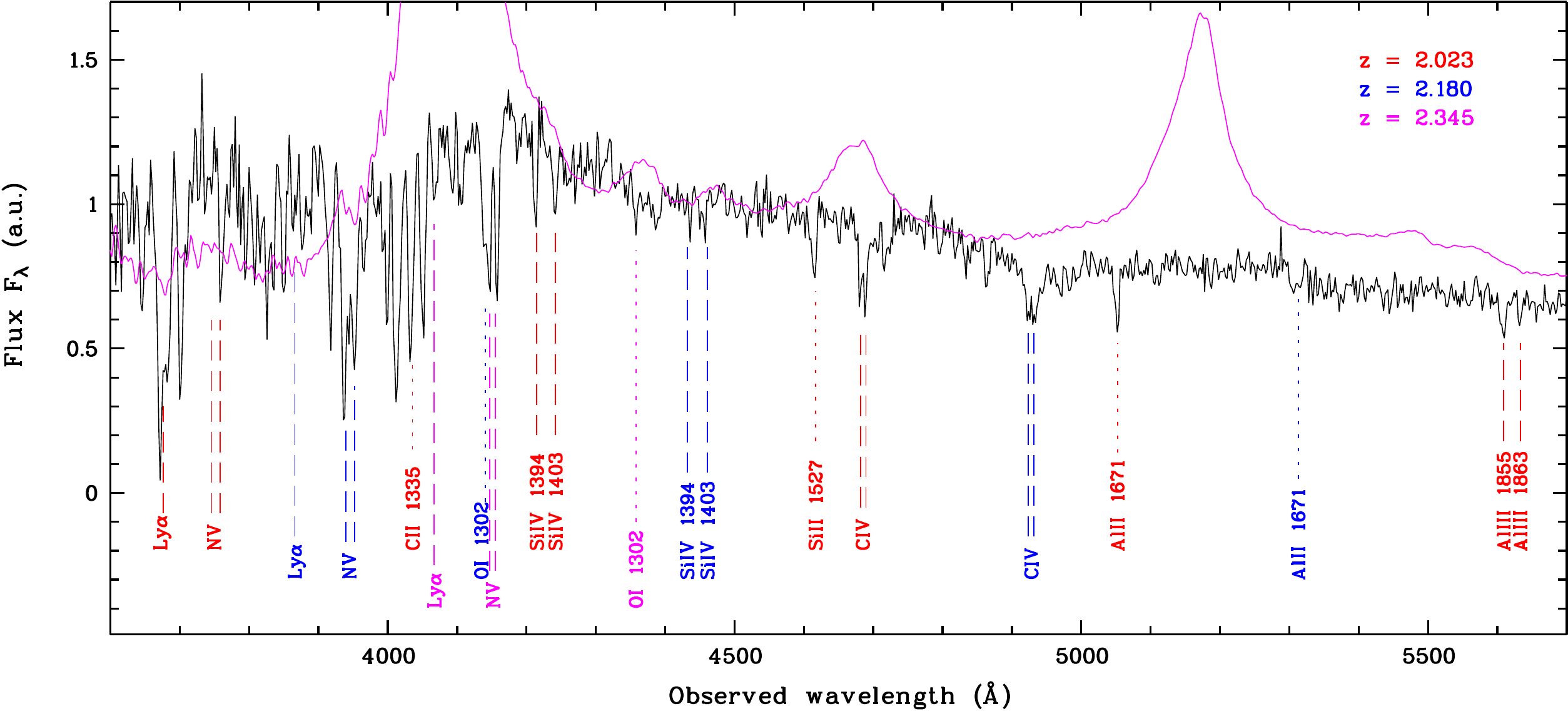} 
\vspace{0.5cm}
\caption{CAFOS B-100 spectrum of VPMS\,J1708+4332 (arbitrary units).
Identified absorption lines, from absorber systems at three different redshifts $z$, are marked by vertical lines with colours corresponding to $z$. 
Dashed vertical lines indicate  common transitions, and dotted lines mark unusual ones. 
NV and CIV denote the line doublets  \ion{N}{v}\,$\lambda\lambda\,1238.8,1242.8$ and \ion{C}{iv}\,$\lambda\lambda\,1548.2,1550.8$, respectively.
For comparison, the redshifted ($z = 2.345$)  and arbitrarily scaled SDSS quasar composite spectrum \citep{Vanden_Berk_2001} has been over plotted (magenta).
}
\label{fig:B100}
\end{center}
\end{figure*}

Nevertheless, we decided to check the EDR3 proper motion of VPMS\,J1708+4332  with its previous position measurements made over a longer time baseline before Gaia started to work. The software  of \citet{Gudehus_2001} allows for a weighted proper motion solution (with the parallax set to zero), using positional data of different quality.  
Using it, we combined the very precise positions from three Gaia data releases, DR1 \citep{Gaia_2016}, DR2 \citep{Gaia_2018}, 
and EDR3  at epochs 2015.0, 2015.5, and 2016.0, with the less precise positions from the best optical catalogues of ground-based observations. 
Two earlier epochs, from 2004 and 2005, are given in SDSS DR12 \citep{Alam_2015}, whereas two intermediate 
epochs (both from 2013) are listed in the First U.S. Naval Observatory Robotic Astrometric Telescope Catalog \citep[URAT1;][]{Zacharias_2015} 
and in Pan-STARRS release 1 \citep[PS1;][]{Chambers_2017}. Compared to the given Gaia catalogue errors of our target ($<$0.3\,mas in DR1 
and $<$0.1\,mas in DR2 and EDR3), its URAT1 errors (11\,mas) appeared to be realistic, whereas PS1 (1.9-3.5\,mas) and especially SDSS (2-3\,mas) 
errors seemed to be too small. According to \citet{Tian_2017}, the typical positional precision in PS1 and SDSS is 10\,mas and 25\,mas, respectively. 
Therefore, we assumed these larger PS1 and SDSS positional errors in our weighted proper motion solution. 
The resulting proper motion, $pmRA=-0.04\pm0.18$\,mas/yr, $pmDE=+0.24\pm0.23$\,mas/yr, was not significant and also did not change if even less precise (errors of several 100\,mas) positions measured on old Schmidt plates, with epochs between 1954 and 1992, were included. 
We considered our confirmed zero proper motion to support the zero parallax measured  in EDR3 and hence extragalactic distance of VPMS\,J1708+4332.  
In Gaia DR3 \citep{Gaia_2022}, our target is listed as a quasar candidate with a probability of being a quasar of $> 0.999$ and with $z = 2.445\pm0.014$.

\begin{figure}[bhtp]
\begin{center}
%\vspace{-0.5cm}
\includegraphics[viewport= 20 20 580 800,width=6.4cm,angle=270]{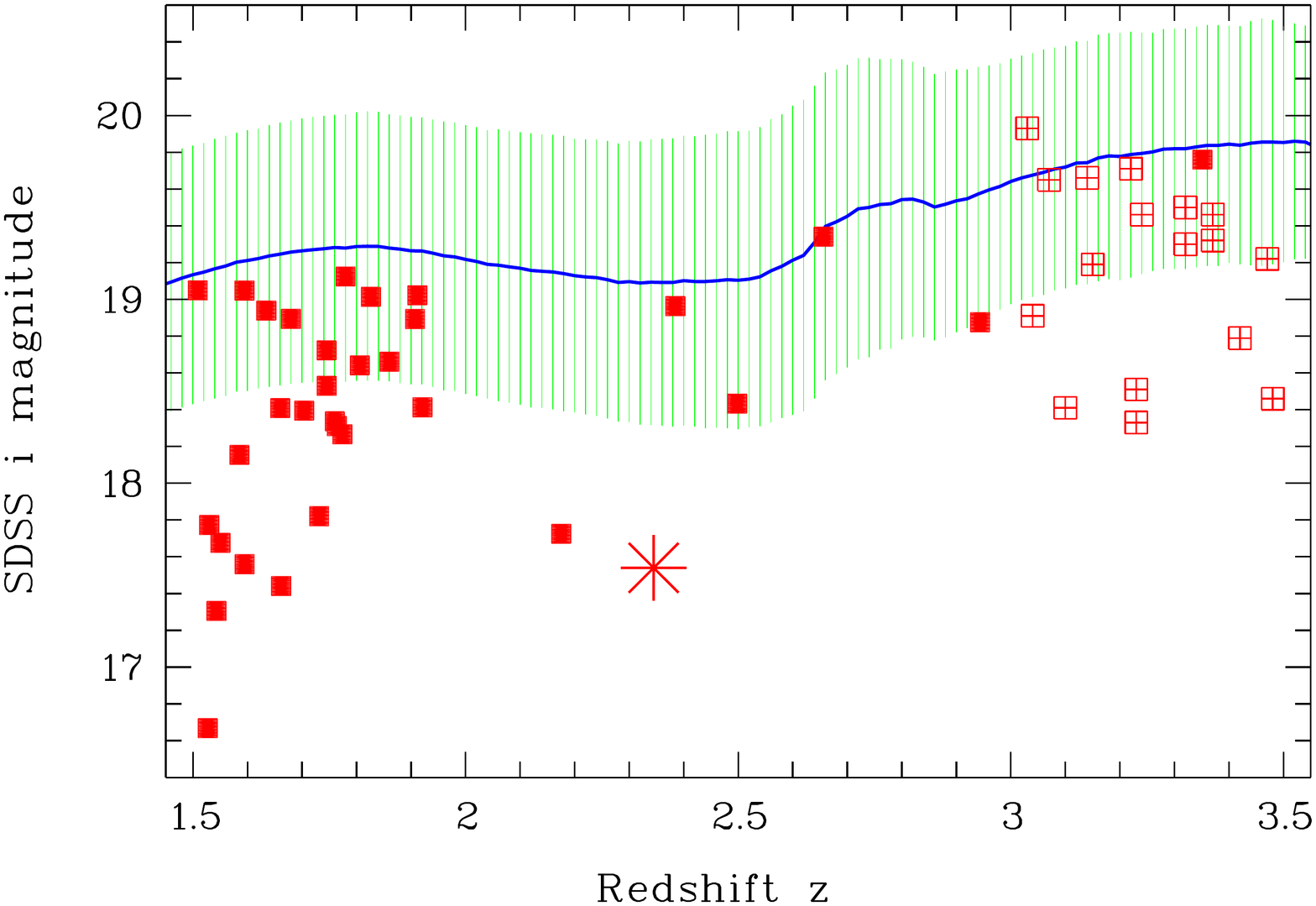} 
\caption{Magnitude-redshift diagram of WLQs for $1.5 \le z \le 3.5$. Red symbols: WLQ sample from \citet{Diamond_2009} (framed crosses); 
EW-selected sub-sample ($\mbox{EW(\ion{C}{iv}})<4.8$\,\AA) from \citet{Meusinger_2014} (squares);
and VPMS\,J1708+4332 (asterisk). For comparison, the distribution of all SDSS DR7 quasars from the \citet{Shen_2011} catalogue is marked by the median relation (thick blue curve) and the standard deviation (hatched green area) .
}
\label{fig:i_z}
\end{center}
\end{figure}

Figure\,\ref{fig:B100} shows the B-100 spectrum of VPMS\,J1708+4332 in the wavelength interval in which narrow absorption lines were found.\footnote{Apart from the slightly lower resolution, the B-200 spectrum (not shown here) looks very similar.} 
For the line identification,  we used the  table of transitions with $\lambda \ge 1215$\,\AA\ seen in BAL quasars \citep{Hall_2002}.
We were able to assign essentially all lines only after we dropped the assumption that they all belong to the same $z$. 
The key to the line identification are the two observed double lines at $\sim 4685$\,\AA\ and 4930\,\AA. 
The wavelength ratio of their two line components is close to that of the  \ion{C}{iv}\,$\lambda\lambda\,1548.2,1550.8$ doublet. 
The identification of these two double lines with \ion{C}{iv} results in $z_{\rm abs,1} = 2.023$ and $z_{\rm abs,2} = 2.128$. 
At these redshifts, several other observed lines could be assigned to common lines from the input table, particularly 
Ly\,$\alpha$, VPMSJ1708
\ion{N}{v}\,$\lambda\lambda\,1239,1243$, 
\ion{Si}{iv}\,$\lambda\lambda\,1394,1403$,
and
\ion{Al}{iii}\,$\lambda\lambda\,1855, 1863$.
Finally, we assume a third absorber at $z_{\rm abs,3} = 2.345$, which explains the strong absorption feature at $\lambda \sim 4150$\,\AA\ as being due to the \ion{N}{v} doublet 
and  the unidentified lines at $\lambda \la 4050$\,\AA\ as being due to the Ly\,$\alpha$ forest. 
The redshift $z_{\rm abs,3}$ is close to  $z$ from the continuum fit (Fig.\,\ref{fig:B400}). 
We simply set $z_{\rm abs,3}$ equal to the systemic redshift $z $, whose correct value can probably best be determined by measuring the $[\ion{O}{III}]\lambda\lambda 4959,5007$ lines by IR spectroscopy in the H-band.

We would like to note that WLQs tend to be relatively bright compared to the typical SDSS quasars at the same redshift \citep[Fig.\,\ref{fig:i_z}; see also][]{Meusinger_2014, Luo_2015}.  
With an i-band apparent magnitude of  $i = 17.54$, VPMS\,J1708+4332 is one of the WLQs with the strongest excess compared to the median relation.

%--------------------------------------------------------------------------------------------------------------------------

\section{Discussion}\label{sect:Discussion}

%--------------------------------------------------------------ght -------------------------------------------------------

The spectrum of  VPMS\,J1708+4332 is comparable to those of the two WLQs SDSS\,J114153.34+021924.3 and SDSS\,J123743.08+630144.9 discussed by \citet{Shemmer_2010}.
With their exceptionally weak, and actually undetectable, UV emission lines, these objects are placed at the low-end tail of the EW distribution of broad emission lines, not only of the type 1 quasars, but also of the WLQs.  

\begin{figure}[htbp]
\begin{center}
\includegraphics[viewport= 20 20 590 800,width=6.5cm,angle=270]{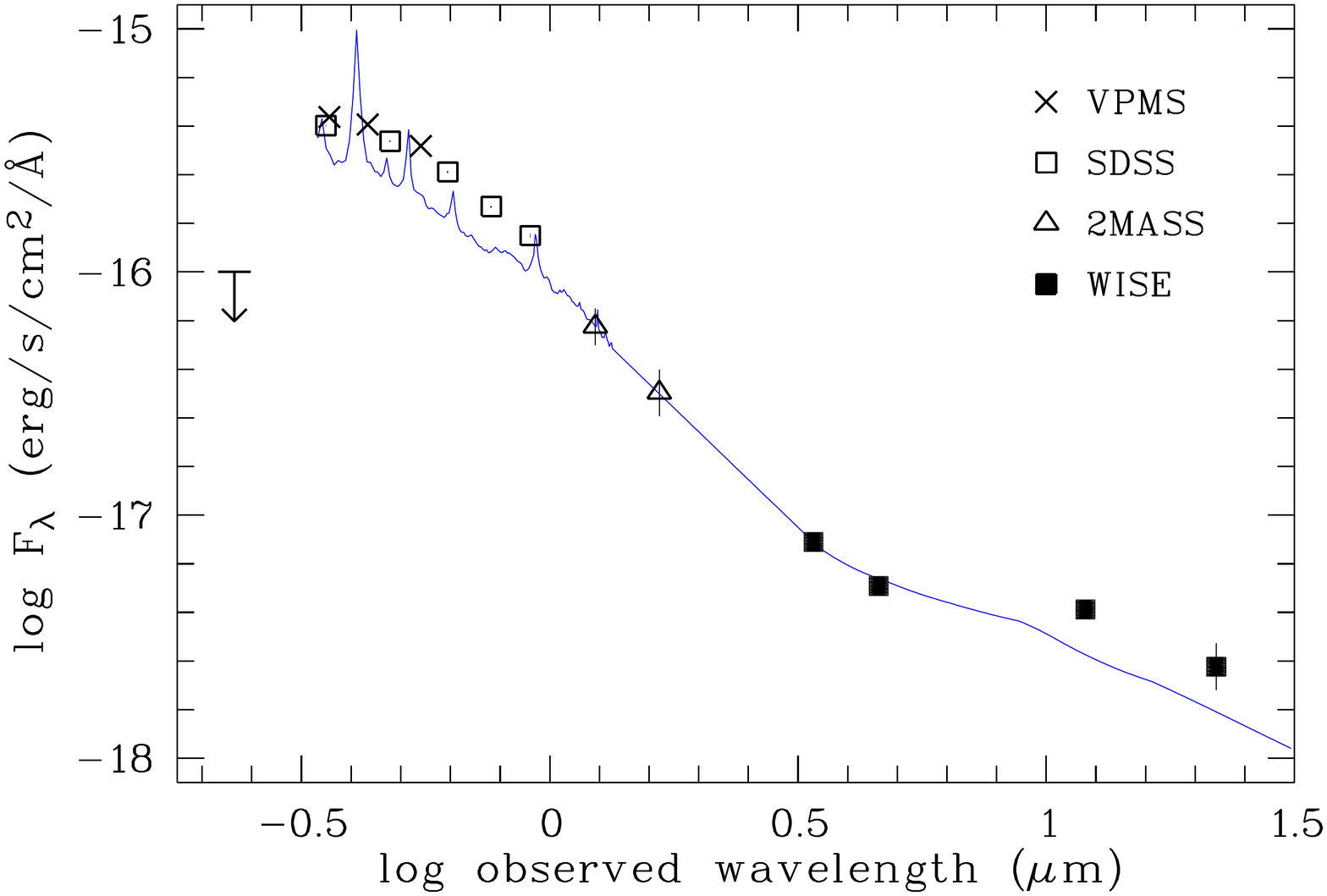} 
\caption{The SED of VPMS\,J1708+4332 based on reddening corrected photometric data from  VPMS, SDSS, 2MASS, and WISE (symbols). 
The error bars are usually smaller than the symbol size.  The downward error is an upper limit from Galex in the near-UV. 
The SWIRE template for bright quasars \citep{Polletta_2007} has been over plotted (blue),  redshifted into the observer frame, and scaled to the observed SED in the near-IR. 
}
\label{fig:SED}
\end{center}
\end{figure}

VPMS\,J1708+4332 is also noteworthy in relation to the slope of the SED.
In general, WLQs exhibit SEDs that are  broadly consistent with the continuum of  normal quasars \citep[e.g.][]{Diamond_2009}.
Figure\,\ref{fig:SED} shows the observed SED of  VPMS\,J1708+4332  based on photometric data from UV to mid-IR. 
In addition to the magnitudes $U, B, V$ from VPMS, $u, g, r, i, z$ from SDSS, and $W1, W2, W3, W4$ from WISE,  the magnitudes $J$ and $H$ from the Two-Micron All-Sky Survey \citep[2MASS;][]{Skrutskie_2006} were used and an upper limit in the near-UV band of the Galaxy
Evolution Explorer \citep[Galex;][]{Morrissey_2007}.
A trend towards a steeper continuum of WLQs at rest-frame wavelengths $\ga 1500$\,\AA\ was noticed by \citet{Meusinger_2014}.
VPMS\,J1708+4332 appears extreme in this respect  (Fig.\,\ref{fig:SED}). 
The spectral slope between  2200\,\AA\ and 4000\,\AA\ (rest frame)  is  $\alpha = -2.31$  
($F_\lambda \propto \lambda^\alpha$). 
For comparison, we estimated a mean slope $\bar{\alpha} = -1.74 \pm 0.35$  for the same wavelength interval from the SDSS spectra of the 45 WLQs from \citet{Meusinger_2014} with $0.7 \le z \le 2.2$ and an equivalent width of the broad \ion{Mg}{ii}\,$\lambda\,2800$ line $< 15$\,\AA.
We note that the fluxes from the SDSS magnitudes agree well with those from VPMS. Because the VPMS fluxes are averaged over five decades, it is  unlikely that the steep increase from near-IR to optical (observer frame)  is due to flux variations in the rest-frame UV.

In the standard model \citep{Shakura_1973}, the radial temperature profile of the AD is $T(r) = T_\ast\,f(r)$, 
where $f(r)$ describes the radial dependence. The temperature parameter $T_\ast$ (in K) is given by 
$T_\ast  = 2.2\times 10^9\,  M_\bullet^{-1/2}\,\dot{M}_\bullet^{1/4}$, where $M_\bullet$ is the black hole (BH) mass in $M_\odot$ and $\dot{M}_\bullet$ is the mass accretion rate in $M_\odot$\,yr$^{-1}$ \citep[e.g.][]{Pereyra_2006}.
A steep slope of the SED in the UV indicates a high  $T_\ast$ and thus,  at a given mass, a high accretion rate.
Because  it is not possible to determine the virial BH mass from the available spectra,  we tentatively estimated $M_\bullet$ and $\dot{M}_\bullet$  from $T_\ast$  and the luminosity $L$.  The combination of the above equation for $T_\ast$ with the relation between $\dot{M}_\bullet$, $M_\bullet$, and  the optical luminosity from \citet[][their Eq. 8]{Davis_2011} results in 
$M_{\bullet} = 2.9\times 10^9\,(L_{\rm opt,45}/\cos \theta)^{0.52}\,T_\ast^{-0.25}$ 
and  
$ \dot{M}_\bullet = 0.13\,(L_{\rm opt,45}/\cos \theta)^{1.04}\,T_\ast^{0.22}$,
where  $L_{\rm opt,45}$ is the optical luminosity $\lambda L_\lambda$ at 4681\,\AA\  (rest frame) in units of $10^{45}$ erg\,s$^{-1}$ and $\theta$ is the AD inclination angle.
The effect of the latter is primarily due to the $\cos \theta$ dependence of the projected AD area.
We have applied the \citet{Davis_2011} relation here for two reasons. Firstly, because the optical luminosity is relatively independent of the 
inner regions of the disk affected by strong relativistic effects.  Secondly, it allows us to compare the results with the WLQ sample of \citet{Meusinger_2014}.
We estimate $T_\ast  =  (1.3\pm0.1)\times 10^5$\,K  from the  $T_\ast - \alpha$ relation for the multi-temperature black-body model of the AD and  $L_{\rm opt} = (2.4 \pm 0.6)\times 10^{46}$\,erg\,s$^{-1}$  from the observed flux density (Fig.\,\ref{fig:SED}). 
This yields $M_{\bullet} = 9.1\times 10^8\,M_\odot$ and  $\dot{M}_\bullet = 57\,M_\odot  \,\mbox{yr}^{-1}$ for a mean disk inclination $\cos \theta = 0.8$.
For the range $\theta = 0\degr ... 60\degr$ usually assumed for quasars, one finds $M_{\bullet} = (8.1 ... 11.6)\times 10^8\,M_\odot$ and  $\dot{M}_\bullet = 45 ... 94\,M_\odot  \,\mbox{yr}^{-1}$ (higher values at larger $\theta$). 
The uncertainties in $L_{\rm opt},  T_\ast,$ and $z$ result in uncertainties of 14\% for $M_\bullet$ and  26\% for $\dot{M}_\bullet$.

The accretion state of quasars is usually expressed by the Eddington ratio $\varepsilon = L_{\rm bol}/L_{\rm Edd}$, where the Eddington luminosity $L_{\rm Edd}$ is determined by $M_\bullet$.  We calculated the bolometric luminosity $L_{\rm bol}$  from the monochromatic luminosities $L_\lambda$ at 
1450\,\AA\ and 3000\,\AA, respectively, using  bolometric corrections $\zeta_{1450}$  and $\zeta_{\rm 3000}$ listed by  \citet{Runnoe_2012}. 
The mean value of the resulting Eddington ratio for $\cos \theta=0.8$ (Table\,\ref{tab:epsilon}) is 
$\epsilon = 2.0$, with an estimated uncertainty of 70\%. For $\theta = 60\degr$ and $0\degr$, we find $\epsilon = 1.66$ and 2.25, respectively. 
VPMS\,J1708+4332  thus belongs to the upper end of the Eddington ratio distribution of both WLQs \citep{Shemmer_2015, Meusinger_2014} 
and type 1 quasars \citep[][]{Kelly_2013}.

\begin{table}[htbp]
\caption{Eddington ratio $\varepsilon$ for $\cos \theta=0.8$ and bolometric corrections $\zeta$ at 1450\,\AA\ and  3000\,\AA\ from different sources.}
\begin{tabular}{lcc} 
\hline\hline 
\noalign{\smallskip}
Reference                    &    $\varepsilon\,(\zeta_{1450})$  &    $\varepsilon\,(\zeta_{3000})$  \\
\hline 
\noalign{\smallskip}
\citet{Elvis_1994}\tablefootmark{a}\tablefootmark{b}         &     1.92                &     1.54 \\
\citet{Richards_2006}\tablefootmark{a}\tablefootmark{b}       &     1.42                &     1.25 \\
\citet{Nemmen_2010}\tablefootmark{a}  &     1.83                &     2.38 \\
\citet{Runnoe_2012}\tablefootmark{a}        &     2.56                &     2.09 \\
\citet{Runnoe_2012}\tablefootmark{c}  &      2.06               &      3.32 \\
\hline                                    
\end{tabular}      
\tablefoot{
\tablefoottext{a}{Taken from Runnoe et al. (2012), their table 2.}
\tablefoottext{b}{Recalculated by Runnoe et al. (2012),}
\tablefoottext{b}{and from their eq. 9.}
}
\label{tab:epsilon}                    
\end{table}

A more general approach also involves the BH spin parameter $a$. \citet{Campitiello_2018} presented analytic approximations of the AD emission for rotating BHs.
Their $L(a,\theta)-\dot{M}_\bullet$ relation  (their Eq.\,B3) confirms a high accretion rate ($\sim 30 M_\odot\,\mbox{yr}^{-1}$ for $a \ga 0.9$ and $60 ... 110 M_\odot\,\mbox{yr}^{-1}$ for $a \approx 0.1$)   for all $\theta \le 60\degr$.
In particular,  they describe the dependence of the peak frequency and the peak luminosity of the AD spectrum on  
$M_{\bullet}, \dot{M}_\bullet, \theta$, and $a$ for  $\theta = 0\degr$. 
We used their Eqs. 12-14  to calculate  $M_{\bullet}$ and  $\dot{M}_\bullet$ as a function of $a$, assuming that the observed peak of the SED (Fig.\,\ref{fig:SED})
is not primarily caused by the Ly$\alpha$ forest. 
The result is  $(M_\bullet [10^9 M_\odot], \dot{M}_\bullet [M_\odot\,\mbox{yr}^{-1}]) = (4.0, 36)$  for $a = 0$  and  $(4.6, 12)$  for $a = 0.998$.
There is no evidence for a pole-on view of VPMS\,J1708+4332, but the effect of the inclination is only moderate for $\theta \la 60\degr$  \citep[e.g. a factor $< 2$ for $L$; see][their Figs. B.1 and B.2]{Campitiello_2018}. Therefore, we can thus assume that this result reflects the correct order of magnitude and does not contradict the above conclusion of a high Eddington ratio.   It should be mentioned that $\epsilon \ga 0.3$ violates the limit for a thin disk and could thus indicate a  slim or thick AD \citep[][]{Abramowicz_1988}, whose vertical structure must be taken into account by numerical models.  Such an inner AD structure has been suggested to be related to a gas component that may shield the broad line region from the ionising continuum \citep{Luo_2015}.

Another remarkable property of VPMS\,J1708+4332 is the blueshift of the two narrow absorption line systems at $z_{\rm abs,1}$ and $z_{\rm abs,2}$.
If they are intrinsic, that is not from intervening systems, the wavelength shifts relative to the adopted systemic redshift  indicate outflows from  the central engine with velocities of $v \ga 0.05c$  and $\ga 0.1c$, respectively.\footnote{$v/c = (R^2-1)/(R^2+1)$, where $R =  (1+ z)/(1 + z_{\rm abs})$}

%--------------------------------------------------------------------------------------------------------------------------

\section{Conclusions}\label{sect:Summary}

%--------------------------------------------------------------------------------------------------------------------------

Supported by the data from Gaia DR3, we have identified VPMS\,J1708+4332  as a WLQ at $z = 2.345$ with undetectable weak UV emission lines.
We tentatively estimate a large BH mass and a high accretion rate close to the Eddington limit, which probably indicates a rapid BH growth  phase.
We identified two narrow absorption line systems with blueshifts which, if intrinsic,  represent outflows with moderately high velocities of  $\sim 0.05c$ and $\sim 0.1c$. This outflow may be  related to the high accretion state.
Near-IR spectroscopy could be helpful to estimate the  virial BH mass based on the H$\beta$ line \citep[e.g.][]{McDowell_1995, Shemmer_2010}. 
VPMS\,J1708+4332  provides another example of unusual AGN that may be under-represented in colour-based quasar surveys.

\begin{acknowledgements}

We thank the referee for  useful comments and suggestions.
We are grateful to the staff of the  Calar Alto observatory for their kind support. 
H. M. acknowledges financial support from the Deutsche Forschungsgemeinschaft under grants Me1359/3 and
Me1350/8.\\

This work has made use of data from the European Space Agency (ESA) mission
{\it Gaia} (\url{https://www.cosmos.esa.int/gaia}), processed by the {\it Gaia}
Data Processing and Analysis Consortium (DPAC,
\url{https://www.cosmos.esa.int/web/gaia/dpac/consortium}). Funding for the DPAC
has been provided by national institutions, in particular the institutions
participating in the {\it Gaia} Multilateral Agreement.

This research has made use of data products from the Sloan
Digital Sky Survey (SDSS). Funding for the SDSS and SDSS-II has been provided by
the Alfred P. Sloan Foundation, the Participating Institutions
(see below), the National Science Foundation, the National
Aeronautics and Space Administration, the U.S. Department
of Energy, the Japanese Monbukagakusho, the Max Planck
Society, and the Higher Education Funding Council for
England. The SDSS Web site is http://www.sdss.org/.
The SDSS is managed by the Astrophysical Research
Consortium (ARC) for the Participating Institutions.
The Participating Institutions are: the American
Museum of Natural History, Astrophysical Institute
Potsdam, University of Basel, University of Cambridge
(Cambridge University), Case Western Reserve University,
the University of Chicago, the Fermi National
Accelerator Laboratory (Fermilab), the Institute
for Advanced Study, the Japan Participation Group,
the Johns Hopkins University, the Joint Institute
for Nuclear Astrophysics, the Kavli Institute for
Particle Astrophysics and Cosmology, the Korean
Scientist Group, the Los Alamos National Laboratory,
the Max-Planck-Institute for Astronomy (MPIA),
the Max-Planck-Institute for Astrophysics (MPA),
the New Mexico State University, the Ohio State
University, the University of Pittsburgh, University
of Portsmouth, Princeton University, the United
States Naval Observatory, and the University of
Washington. \\

This publication has made use of the VizieR catalogue access
tool, CDS, Strasbourg, France, and of the NASA/IPAC Infrared Science Archive (IRSA), operated by the 
Jet Propulsion Laboratories/California Institute of Technology, founded
by the National Aeronautic and Space Administration.
In particular,  
this publication makes use of data products from the Wide-field Infrared Survey Explorer, 
which is a joint project of the University of California, Los Angeles, 
and the Jet Propulsion Laboratory/California Institute of Technology, 
funded by the National Aeronautics and Space Administration. 
In addition, we used data products from the Two Micron All Sky Survey, which is a 
joint project of the University of Massachusetts and the Infrared Processing and
Analysis Center/California Institute of Technology, funded by the National Aeronautics 
and Space Administration and the National Science Foundation.
We also used observations made with the NASA Galaxy Evolution Explorer, GALEX, which is
operated for NASA by the California Institute of Technology under NASA contract NAS5-98034.

\end{acknowledgements}

%% WARNING
%%-------------------------------------------------------------------
%% Please note that we have included the references to the file aa.dem in
%% order to compile it, but we ask you to:
%%
%% - use BibTeX with the regular commands:
%%   \bibliographystyle{aa} % style aa.bst
%%   \bibliography{Yourfile} % your references Yourfile.bib
%%
%% - join the .bib files when you upload your source files
%-------------------------------------------------------------------

\bibliographystyle{aa} % style aa.bst
\bibliography{literature} % your references Yourfile.bib
%\bibliography{lit_temp} % your references Yourfile.bib

\end{document}